\documentclass[a4paper,conference]{IEEEtran}

\IEEEsettopmargin{t}{30mm}
\IEEEquantizetextheight{c}
\IEEEsettextwidth{14mm}{14mm}
\IEEEsetsidemargin{c}{0mm}

\usepackage[utf8]{inputenc}
\usepackage[T1]{fontenc}
\usepackage{url}
\usepackage{ifthen}
\usepackage{cite}
\usepackage[cmex10]{amsmath} 
\usepackage{amssymb}
\usepackage{algorithm}
\usepackage{algpseudocode}
\usepackage{graphicx}
\usepackage{color}
\usepackage{xcolor}
\usepackage{cite}
\usepackage{subcaption}
\usepackage{comment}
\usepackage{mathtools}
\usepackage{tabulary}
\usepackage[nohyperlinks]{acronym}
\usepackage{upgreek}
\usepackage{placeins}
\usepackage[super]{nth}
\usepackage{dsfont}
\usepackage{hyperref}
\usepackage{array}
\usepackage{soul}
\usepackage{dsfont}
\usepackage{cancel}
\usepackage{multirow}
\usepackage{bm}
\usepackage{bbm}
\usepackage{amsthm}
\usepackage{url}
\usepackage[acronym]{glossaries}
\usepackage{balance, flushend}
\glsdisablehyper

\newacronym{GO}{GO}{Goal-Oriented}
\newacronym{RIS}{RIS}{Reconfigurable Intelligent Surface}
\newacronym{NL}{NL}{NonLinear}
\newacronym{SIM}{SIM}{Stacked Intelligent Metasurfaces}
\newacronym{DNN}{DNN}{Deep artificial Neural Network}
\newacronym{NN}{NN}{neural network}
\newacronym{EI}{EI}{Edge Inference}
\newacronym{EM}{EM}{ElectroMagnetic}
\newacronym{TOC}{GOC}{Goal-Oriented Communications}
\newacronym{TX}{TX}{Transmitter}
\newacronym{RX}{RX}{Receiver}
\newacronym{UE}{UE}{User Equipment}
\newacronym{BS}{BS}{Base Station}
\newacronym{LoS}{LoS}{Line-of-Sight}
\newacronym{NLoS}{NLoS}{Non Line-of-Sight}
\newacronym{CNN}{CNN}{Convolutional Neural Network}
\newacronym{FFNN}{FFNN}{Feed-Forward Neural Network}
\newacronym{ReLU}{ReLU}{Rectified Linear Unit}
\newacronym{JSCC}{JSCC}{Joint Source Channel Coding}
\newacronym{E2E}{E2E}{End-to-End}
\newacronym{D2D}{D2D}{Device-to-Device}
\newacronym{DeepSC}{DeepSC}{Deep Semantic Communications}
\newacronym{SISO}{SISO}{Single Input Single Output}
\newacronym{MISO}{MISO}{Multiple Input Single Output}
\newacronym{MIMO}{MIMO}{Multiple-Input Multiple-Output}
\newacronym{OAC}{OAC}{Over-the-Air Computation}
\newacronym{MSE}{MSE}{Mean Squared Error}
\newacronym{SOTA}{SotA}{State of the Art}
\newacronym{SGD}{SGD}{Stochastic Gradient Descent}
\newacronym{IoT}{IoT}{Internet of Things}
\newacronym{AE}{AE}{Auto-Encoder}
\newacronym{PDF}{PDF}{Probability Density Function}
\newacronym{AWGN}{AWGN}{Additive White Gaussian Noise}
\newacronym{CSI}{CSI}{Channel State Information}
\newacronym{CFR}{CFR}{Channel Frequency Response}
\newacronym{ISAC}{ISAC}{Integrated Sensing and Communications}
\newacronym{iid}{i.i.d.}{independent and identically distributed}
\newacronym{MINN}{MINN}{Metasurfaces-Integrated Neural Network}
\newacronym{MS}{MS}{MetaSurface}
\newacronym{CE}{CE}{Cross Entropy}
\newacronym{HRIS}{HRIS}{Hybrid RIS}
\newacronym{RF}{RF}{Radio-Frequency}
\newacronym{VAE}{VAE}{Variational Auto-Encoder}
\newacronym{PSK}{PSK}{Phase Shift Keying}
\newacronym{SVD}{SVD}{Singular Value Decomposition}
\newacronym{MLP}{MLP}{Multi Layer Perceptron}
\newacronym{SNR}{SNR}{Signal-to-Noise Ratio}
\newacronym{MI}{MI}{Mutual Information}
\newacronym{MEC}{MEC}{Multi-access Edge Computing}
\newacronym{WMMSE}{WMMSE}{Weighted Minimum Mean Square Error}
\newacronym{ANN}{ANN}{Artificial Neural Network}
\newacronym{SLFN}{SLFN}{Single hidden Layer Feedforward Network}
\newacronym{ELM}{ELM}{Extreme Learning Machine}
\newacronym{LTI}{LTI}{Linear Time Invariant}
\newacronym{OTA}{OTA}{Over-The-Air}
\newacronym{GOC}{GOC}{Goal-Oriented Communications}
\newacronym{AM}{AM}{Amplitude Modulation}
\newacronym{LS}{LS}{Least Squares}
\newacronym{DC}{DC}{Direct Current}
\newacronym{RFC}{RFC}{Radio-Frequency Chain}
\newacronym{ML}{ML}{Machine Learning}
\newacronym{XL}{XL}{eXtremely Large}
\newacronym{PGD}{PGD}{Projected Gradient Descent}
\newacronym{WBCD}{WBCD}{Wisconsin Breast Cancer Dataset}
\newacronym{DMA}{DMA}{Dynamic Metasurface Antenna}
\newacronym{CMS}{CMS}{Cascaded MetaSurfaces}

\title{Over-The-Air Extreme Learning Machines with\\ XL Reception via Nonlinear Cascaded Metasurfaces\\ \vspace{0.2cm}\normalfont\large \textit{(Invited Presentation at 2026 International Zurich Seminar on Information and Communication)}}

\newcommand\blfootnote[1]{%
  \begingroup
  \renewcommand\thefootnote{}\footnotetext{#1}%
  \addtocounter{footnote}{-1}%
  \endgroup
}

\author{%
  \IEEEauthorblockN{Kyriakos Stylianopoulos\IEEEauthorrefmark{1},
  Mattia Fabiani\IEEEauthorrefmark{2}\IEEEauthorrefmark{3},
  Giulia Torcolacci\IEEEauthorrefmark{2}\IEEEauthorrefmark{3},
  Davide Dardari\IEEEauthorrefmark{2}\IEEEauthorrefmark{3}, George C. Alexandropoulos\IEEEauthorrefmark{1}
  }
  \IEEEauthorblockA{\IEEEauthorrefmark{1}%
  Department of Informatics and Telecommunications, National and Kapodistrian University of Athens, 16122 Athens, Greece} 
  \IEEEauthorblockA{\IEEEauthorrefmark{2}%
  DEI, University of Bologna, 40136 Bologna, Italy}
  \IEEEauthorblockA{\IEEEauthorrefmark{3}%
  National Laboratory of Wireless Communications (WiLab), CNIT, 40136 Bologna, Italy\\
    e-mails: \{kstylianop, alexandg\}@di.uoa.gr, \{mattia.fabiani5, g.torcolacci, davide.dardari\}@unibo.it
}
}

\begin{document}

\maketitle
\blfootnote{This work has been supported by the SNS JU projects 6G-DISAC and TIMES under the EU's Horizon Europe research and innovation program under grant agreement numbers 101139130 and 101096307, respectively. 
The work of G. Torcolacci was funded by an NRRP Ph.D. grant.}

\begin{abstract}
The recently envisioned goal-oriented communications paradigm calls for the application of inference on wirelessly transferred data via Machine Learning (ML) tools. An emerging research direction deals with the realization of inference ML models directly in the physical layer of Multiple-Input Multiple-Output (MIMO) systems, which, however, entails certain significant challenges. In this paper, leveraging the technology of programmable MetaSurfaces (MSs), we present an eXtremely Large (XL) MIMO system that acts as an Extreme Learning Machine (ELM) performing binary classification tasks completely Over-The-Air (OTA), which can be trained in closed form. The proposed system comprises a receiver architecture consisting of densely parallel placed diffractive layers of XL MSs, also known as Stacked Intelligent Metasurfaces (SIM), followed by a single reception radio-frequency chain. The front layer facing the XL MIMO channel consists of identical unit cells of a fixed NonLinear (NL) response, whereas the remaining layers of elements of tunable linear responses are utilized to approximate OTA the trained ELM weights. Our numerical investigations showcase that, in the XL regime of MS elements, the proposed XL-MIMO-ELM system achieves performance comparable to that of digital and idealized ML models across diverse datasets and wireless scenarios, thereby demonstrating the feasibility of embedding OTA learning capabilities into future wireless systems.  
\end{abstract}


\section{Introduction}
Future wireless networks will leverage \gls{EI} to jointly train transceiver pairs as end-to-end \gls{ML} models for efficient sensory data inference~\cite{GO_review2}.
By exchanging task-specific representations through the channel, \gls{EI} overcomes the inefficiencies of conventional decoupled designs in terms of data rate and computational burden, since feature extraction is performed alongside encoding at the \gls{TX}, while the \gls{RX} directly infers the target values instead of reconstructing the input data~\cite{GJZ24_SIM_TOC, Stylianopoulos_GO}.

To further improve computational efficiency, \gls{OTA} computing exploits the wireless propagation domain by performing computation directly through the superposition of traveling \gls{RF} signals~\cite{OTA_review}.
The \gls{OTA} paradigm has recently attracted interest in wireless \gls{ML} applications.
Specifically, architectures based on \glspl{MS} have been proposed to emulate \gls{DNN} layers~\cite{XYN18_Diffractive_DNN, Momeni2022_wave_DL_computing,Stylianopoulos_GO,GJZ24_SIM_TOC} for \gls{OTA} inference, which are trained through backpropagation.
Additionally, another family of approaches uses \gls{MS}-controlled channel responses to approximate digitally trained \gls{DNN} weight (or similar) matrices \gls{OTA}~\cite{AYG24_SIM_DFT, Gunduz_Layer_Approximation, Pandolfo_SIM_semantic_alignment}.
Nevertheless, many existing systems still rely on digital processing and lack theoretical foundations.
Crucially, most \gls{MS}-based \gls{DNN} implementations are only capable of linear operations~\cite{Styl:25}, which drastically reduces the approximation capability of the developed models.

Addressing some of these gaps, \cite{Stylianopoulos_MIMO_ELM} designed an \gls{XL} \gls{MIMO} system performing as an \gls{ELM}~\cite{HuaZhuSie:06} to execute \gls{DNN} operations partially \gls{OTA}, treating the channel as random hidden-layer weights and the \gls{RX} analog combiner as the output layer.
This approach exhibits fast training and re-tuning as the wireless channel evolves and is proven to be a universal function approximator.
However, it faces practical limitations and scalability concerns due to real-valued signal constraints and hardware complexity induced by the use of \gls{NL} power amplifiers and numerous \gls{RF} chains. 

In this paper, capitalizing on the complex-domain \gls{ELM} framework~\cite{LiHuaSarSun:05, HuaLiCheSie:08} and building upon recent \gls{NL} metamaterial advancements~\cite{Styl:25,FabTorDar:25,Omr:25}, we present an \gls{XL}-\gls{MIMO}-\gls{ELM} system with an \gls{RX} structure comprising Cascaded diffractive MSs (CMS), also known as Stacked Intelligent Metasurfaces (SIM)\cite{AXN23_SIM}, followed by a single antenna element and its respective \gls{RF} chain. The initial \gls{MS} layer facing
the \gls{MIMO} channel consists of unit cells of identical fixed \gls{NL} responses and serves as the \gls{ELM}'s activation function, whereas the remaining linear \gls{MS} layers perform trainable \gls{OTA} combining, thereby approximating the digital \gls{ELM} weights. The proposed \gls{NL}-CMS-\gls{ELM} system enables fast training and minimizes digital processing at the \gls{XL} reception side, while significantly reducing the hardware complexity with respect to~\cite{Stylianopoulos_MIMO_ELM}'s \gls{XL}-\gls{MIMO}-\gls{ELM} framework.

\textbf{Notation:} Vectors, matrices, and sets are expressed in lowercase bold, uppercase bold, and uppercase calligraphic typefaces, respectively.
$\mathbf{X}^\top$ and $\mathbf{X}^{\rm H}$  denote the transpose and conjugate transpose of $\mathbf{X}$.
$[\mathbf{x}]_i$ is used to denote the $i$th element of $\mathbf{x}$.
$|\mathcal{X}|$ represents the cardinality of the set $\mathcal{X}$, $||\mathbf{X}||_{\rm F}$ denotes the Frobenius norm of $\mathbf{X}$, ${\rm diag(\mathbf{x})}$ creates a square matrix with the elements of $\mathbf{x}$ placed along its main diagonal.
$\{\cdot\}$ expresses a set or collection, while $\mathbbm{1}_{\texttt{cond}}$ is the indicator function taking value $1$ if the condition $\texttt{cond}$ holds, and $0$ otherwise.
Finally, $\mathbf{X} \sim \mathcal{CN}(\mathbf{0}, \sigma^2\mathbf{I})$ signifies that $\mathbf{X}$ follows the complex standard Gaussian distribution with variance $\sigma^2$, $\jmath\triangleq\sqrt{-1}$ is the imaginaty unit, and $z \in \mathbb{C}$, which can be written as $z \triangleq \mathfrak{Re}(z) + \jmath \mathfrak{Im}(z) = |z| \exp(\jmath~ {\rm arg}\{z\})$.

\section{The Proposed \gls{XL} \gls{MIMO} System Model}\label{sec:system-model}

\begin{figure}
    \centering
    \includegraphics[width=1.0\linewidth]{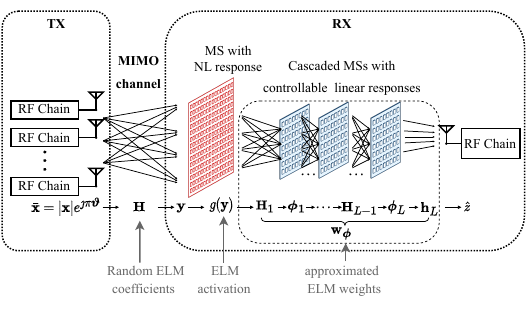}
    \caption{The proposed \gls{XL} \gls{MIMO} system for implementing the proposed \gls{NL}-\acrshort{CMS}-\gls{ELM} framework. Channel and \gls{MS} responses are used as components of the \gls{ELM} algorithm to realize \gls{OTA} inference. The flow of computational during the forward pass is also sketched.}
    \label{fig:system-model}
\end{figure}

Consider a narrowband \gls{XL} \gls{MIMO} system comprising an $N_{\rm t}$-antenna \gls{TX} and an \gls{RX} consisting of multiple \gls{XL} diffractive \gls{MS} layers with different functionalities, whose outputs are cascaded and ultimately collected by a single reception \gls{RF} chain, as depicted in Fig.\ref{fig:system-model}.
Instead of performing conventional wireless communications, the system is trained end-to-end to act as an \gls{OTA} function approximator~\cite{Cyb:89, Stylianopoulos_MIMO_ELM}.
In particular, given an offline dataset $\mathcal{D} \triangleq \{(\mathbf{x}^{(i)}, z^{(i)})\}_{i=1}^D$ of $D$ input-target pairs, the \gls{XL} \gls{MIMO} system is designed to approximate the $\mathbf{x} \to z$ mapping so that the \gls{TX} observes the input data $\mathbf{x}$ (not necessarily belonging to $\mathcal{D}$) and the \gls{RX} estimates its (unobserved) target value $z$.
From this perspective, the system is intended to perform \gls{GOC}~\cite{Goldreich_GoalOriented_theory}, with all computational processing performed exclusively \gls{OTA} in the \gls{RF} domain.

We assume that $\mathbf{x}^{(i)} \in [0,1]^{N_{\rm t}}$ and $z^{(i)} \in \{-1,+1\}$, i.e., the dimension of the data observations are equal to the number of \gls{TX} antennas and the dataset is used for binary classification. Note that \gls{TX} may introduce its own trainable feature extraction model to reduce the dimensionality of the $\mathbf{x}$~\cite{Stylianopoulos_GO, GJZ24_SIM_TOC}; this is left for future work.
For reasons that will be described later, the data points are subject to an \gls{AM}, hence, the transmitted signal is given by the expression:
\begin{equation}\label{eq:transmit-signal}
    \bar{\mathbf{x}} \triangleq |\mathbf{x}| \exp(\jmath \pi  \bm \vartheta) \in \mathbb{C}^{N_{\rm t} \times 1},
\end{equation}
where the elements of $\bm \vartheta$ may be chosen arbitrarily.
The baseband representation of the signal arriving at the first (front) \gls{MS} layer of the \gls{RX}, which is composed of $N_{\rm r} \triangleq N^{\rm hor}_{\rm r}\times N^{\rm vert}_{\rm r}$ metamaterial elements, can be expressed as follows:
\begin{equation}\label{eq:received-signal}
    \mathbf{y} \triangleq \mathbf{H} \bar{\mathbf{x}} \in \mathbb{C}^{N_{\rm r} \times 1},
\end{equation}
where $\mathbf{H} \in \mathbb{C}^{N_{\rm r} \times N_{\rm t}}$ represent the \gls{XL} \gls{MIMO} channel response.
For reasons discussed later, we consider the case where $\mathbf{H}$ follows the Rayleigh fading distribution and remains quasi-static for the duration of the training process.

\subsection{Proposed \gls{RX} Architecture}
We propose an \gls{RX} architecture consisting of densely parallel  placed diffractive \gls{MS} layers forming a CMS architecture whose purpose is to transform the impinging wave towards a single antenna element attached to a reception \gls{RF} chain, as it will be explained in the following.

\subsubsection{Front \gls{MS} Layer with \gls{NL} Activation}
We consider the front diffractive \gls{MS} layer composed of unit elements applying a memoryless \gls{NL} transformation. Denoting with $F(\cdot)$ the bandpass response of the generic element of the \gls{NL} \gls{MS}, the baseband-equivalent output $g(\cdot)$ preserves the phase of the input while transforming its envelope through the first-order harmonic extraction.  The resulting element-wise mapping of the \gls{MS} is expressed as $g(\mathbf{y}) \triangleq C(|\mathbf{y}|) \exp(\jmath {\rm arg}\{\mathbf{y}\})$, where $C(\cdot)$ denotes the \gls{AM}/\gls{AM} characteristic derived as \cite{FabTorDar:25}:
\begin{equation} \label{eq:C}
C(v)=\frac{2}{\pi} \int_{0}^{\pi} F(v\cos(\phi)) \cos(\phi) \text{d}\phi.
\end{equation}

In this paper, we consider an element-wise thresholding device characterized overall by the positive bias $\mathbf{b} \in \mathbb{R}^{N_{\rm r} \times 1}_+$, whose elements are drawn from an appropriate distribution during fabrication, hence, ensuring low complexity. From \eqref{eq:C}, the transform yields the following piecewise mapping:
\begin{equation}
    C(|\mathbf{y}|) = \begin{cases} 
    \mathbf{0}, & |\mathbf{y}| \le \mathbf{b} \\
    \begin{aligned}
    \frac{1}{\pi} \left( |\mathbf{y}| \arccos\left(\frac{\mathbf{b}}{|\mathbf{y}|}\right)\right. & \\
    \left.- \mathbf{b} \sqrt{1 - \left(\frac{\mathbf{b}}{|\mathbf{y}|}\right)^2}\right) & ,
    \end{aligned} & |\mathbf{y}| > \mathbf{b}
    \end{cases}
    \label{eq: C[|y|]}
\end{equation}
where all operations are applied element-wise. While this expression captures the exact physical behavior of the \gls{MS} elements, the transcendental terms are computationally demanding for \gls{ELM} training. 
By approximating the transition for $|\mathbf{y}| > \mathbf{b}$ as a linear slope, $g(\mathbf{y})$ implements a rudimentary magnitude-dependent \gls{ReLU} activation function with the inclusion of the bias term. 
Consequently, the activation function for our system model is approximated as:
\begin{equation}\label{eq:activation-function}
    g(\mathbf{y}) \simeq \frac{1}{2}\max\left(\mathbf{0}, |\mathbf{y}| - \mathbf{b}\right) \exp\left(j {\rm arg}\{\mathbf{y}\}\right),
\end{equation}
where $\mathbf{b}$ serves as the effective hardware-induced threshold vector that facilitates \gls{NL} processing in the \gls{RF} domain. In \cite{FabTorDar:25}, a possible practical implementation based on diodes is proposed.
We note that an analytical activation named ``modReLU``, closely resembling~\eqref{eq:activation-function}, has been used in the context of complex-valued~\glspl{DNN} in~\cite{Arj16_ComplexRNN} and follow-up works.

\subsubsection{Cascade of MSs with Trainable Linear Responses}
The outputs of the front diffractive \gls{NL} \gls{MS} layer are then passed to a cascade of $L$ diffractive linear \glspl{MS}.
Each layer comprises a square grid of $N_l$ elements ($l=1,2, \ldots,L$) spaced by $\lambda/ 2$, with $\lambda = c/f_0$ denoting the wavelength at the carrier frequency $f_0$, and $c$ is the speed of light. Let $\mathbf{H}_l \in \mathbb{C}^{N_l \times N_{l-1}}$ denote the signal propagation coefficients between the $(l-1)$-th and the $l$-th \gls{MS} layer, where $N_0 = N_{\rm r}$ indicates the number of elements of CMS's front \gls{MS} layer, and $\mathbf{h}_L \in \mathbb{C}^{N_L \times 1}$ represent the propagation between the last \gls{MS} layer and the single antenna element attached to the \gls{RX} \gls{RF} chain. Typical works utilizing the technology of stacked intelligent metasurfaces (SIM)~\cite{AXN23_SIM} assume free-space propagation between the \gls{MS} layers, i.e., considering an anechoic enclosure, and model the element-to-element propagation through geometric optics~\cite{AXN23_SIM, AYG24_SIM_DFT, XYN18_Diffractive_DNN}.
In this work, we model each $\mathbf{H}_l$ as a full rank pseudo-random matrix, which implies a reverberating enclosure around the layers~\cite{PhysFad, Tacid_PhysFad}.
Arguably, this choice is more realistic as it accounts for multipath components arising from imperfections of the enclosure and allows for the \gls{MS} layers to be placed arbitrarily close.
Moreover, the richer propagation diversity compared to geometric optics provides substantial gains for the optimization framework, as explained in the following section.

The responses of each $l$-th \gls{MS} layer are expressed as ${\bm \phi}_l \triangleq {\bm \alpha}_l \exp(\jmath \pi {\bm \theta}_l)$, with the amplitudes ${\bm \alpha}_l \in [0,1]^{N_l \times 1}$ and the phase shifts ${\bm \theta}_l \in [0,1]^{N_l \times 1}$ being {\em controllable} parameters.
By defining ${\bm \Phi}_l \triangleq {\rm diag}({\bm \phi}_l)$ and ${\bm \varphi} \triangleq \{{\bm \phi}_l \}_{l=L-1}^1$, the overall transfer function of the $L$ cascaded linear \glspl{MS} is given by:
\begin{equation}\label{eq:sim-response}
    \mathbf{w}_{\bm \varphi} \triangleq \mathbf{h}^\top_L \prod_{l=L-1}^{1} {\bm \Phi}_{l} \mathbf{H}_{l} \in \mathbb{C}^{1 \times N_{\rm r}}.
\end{equation}
Consequently, the signal at the output of the \gls{RX} \gls{RF} chain is:
\begin{equation}\label{eq:output-signal}
    \hat{z} \triangleq \mathbf{w}_{\rm \varphi} \, g(\mathbf{y}) + \tilde{n},
\end{equation}
where $\tilde{n}\sim\mathcal{CN}(0,\sigma^2)$ represents the \gls{AWGN}.

\section{Training as an Extreme Learning Machine}
The previously presented \gls{XL} \gls{MIMO} system may be regarded as an affine transformation of the input (through expression~\eqref{eq:received-signal} and the bias term in~\eqref{eq:activation-function}) followed by the \gls{NL} activation inside~\eqref{eq:activation-function} and the final (linear) weighted sum through the cascaded response of the $K$ linear \gls{MS} layers described in~\eqref{eq:output-signal}.
Subsequently, the performed computations are equivalent to those of a single-hidden-layer feedforward neural network, which motivates its deployment as a function approximator.
Nevertheless, not all components are controllable.
The channel responses, $\mathbf{H}$ and $\mathbf{H}_l$'s, and the activation biases $\mathbf{b}$, in particular, are treated as random coefficients.
In that regard, we leverage the \gls{ELM} mathematical framework~\cite{HuaZhuSie:06, LiHuaSarSun:05, Stylianopoulos_MIMO_ELM} used for \gls{ML} inference, which allows random parameters alongside trainable weights, enabling both the training procedure and its theoretical guarantees to be rigorously described.

In approximating the $\mathbf{x} \to z$ mapping based on $\mathcal{D}$, we define the \gls{ELM} activation matrix $\mathbf{G} \in \mathbb{C}^{D \times N_{\rm r}}$ as the transpose of the activated signals at the \gls{RX}'s front \gls{MS} layer:
\begin{equation}\label{eq:hidden-layer-outputs}
    \mathbf{G} \triangleq [g(\mathbf{y}^1), \dots, g(\mathbf{y}^D)]^{\top}.
\end{equation}
By further denoting the vector of target values for the whole dataset as $\mathbf{z} \triangleq [z^1 , \dots , z^D]^\top \in \mathbb{C}^{D \times 1}$,
the cascaded response of the $L$ linear \glspl{MS}, $\mathbf{w}_{\bm \varphi}$, can be optimized to minimize the \gls{LS} error between the target and output values, following the standard \gls{ELM} formulation, as follows:
\begin{equation}\label{eq:LS-objective}
    \mathbf{w}^{*} \triangleq \arg \min_{\mathbf{w}_{\bm \varphi}} \| \mathbf{z} - \mathbf{G} \mathbf{w}_{\bm \varphi}\|_{2}^{2}.
\end{equation}
This yields the closed-form solution:
\begin{equation} \label{eq:weights-opt}
    \mathbf{w}^{*} = \left(\mathbf{G}^{\rm H} \mathbf{G} + \ell \mathbf{I}\right)^{-1}\mathbf{G}^{\rm H} \mathbf{z} \in \mathbb{C}^{N_{\rm r} \times 1},
\end{equation}
which accounts for Tikhonov regularization, controlled by the hyperparameter $\ell>0$, to ensure generalization beyond the training dataset $\mathcal{D}$. 

While~\eqref{eq:weights-opt} provides a fast and convenient way to determine the cascaded \gls{MS} response, $\mathbf{w}_{\bm \varphi}$ is not directly controllable because of the physical limitations introduced by the diffractive \glspl{MS}. Instead, the trainable parameters are ${\bm a}_l$'s and ${\bm \theta}_l$'s of ${\bm \varphi}$.
Therefore, as the second step of our training approach, we take inspiration from works that use \gls{MS} responses to approximate arbitrary matrices, including \gls{DNN} weights~\cite{AYG24_SIM_DFT, Gunduz_Layer_Approximation, Pandolfo_SIM_semantic_alignment}, and we choose to find appropriate ${\bm \varphi}$ values that approximate $\mathbf{w}^\ast$ as:
\begin{align}\label{eq:weight-approximation-obj}
    {\bm \varphi}^\ast = \{ {\bm \phi}^\ast_l \}_{l=1}^L \triangleq {\rm arg}\,\min_{\mathbf{w}_{\bm \varphi},\rho}\left\| \mathbf{w}^\ast -  \rho \mathbf{w}_{\bm \varphi}^\top \right\|_{\rm F},
\end{align}
where $\rho > 0$ is a scaling term to compensate for the inadvertent signal attenuation induced by~\eqref{eq:sim-response}.
In practice, this implies the inclusion of dedicated amplification at the single reception \gls{RF} chain (e.g., via the low noise amplifier).
Since~\eqref{eq:sim-response} provides an analytic expression for $\mathbf{w}_{\bm \varphi}$,~\eqref{eq:weight-approximation-obj} is differentiable with respect to ${\bm a}_l$ and ${\bm \theta}_l$. Consequently, the minimization is performed via \gls{PGD} by projecting ${\bm a}_l$, ${\bm \theta}_l$, and $\rho$ onto their respective feasible sets.
The overall training procedure is summarized in Algorithm~\ref{alg:training}, while the use of the trained system for inferring a target value given an input data point is detailed in Algorithm~\ref{alg:inference}.

\begin{algorithm}[t]
\caption{Training of the Proposed \gls{NL}-CMS-\gls{ELM}}
\label{alg:training}
\begin{algorithmic}[1]
\State \textbf{Inputs:} Dataset $\mathcal{D} = \{(\mathbf{x}^{(i)}, z^{(i)})\}_{i=1}^D$, \gls{TX}-\gls{RX} channel $\mathbf{H}$, \gls{MS} propagation coefficients $\{\mathbf{H}_l\}_{l=1}^L$ and $\mathbf{h}_L$.
\For{$i=1,\dots,D$}
\State Construct transmit signal $\bar{\mathbf{x}}^{(i)}$ via \eqref{eq:transmit-signal}.
\State Transmit $\bar{\mathbf{x}}^{(i)}$ to obtain $\mathbf{y}^{(i)}$ at front \gls{MS} layer via \eqref{eq:received-signal}.
\State Obtain $g(\mathbf{y}^{(i)})$ via the activation of \eqref{eq:activation-function}.
\EndFor
\State Collect all activations in $\mathbf{G}$ via \eqref{eq:hidden-layer-outputs}.
\State Compute ideal weights $\mathbf{w}^\ast$ via \eqref{eq:weights-opt}.
\State Apply \gls{PGD} on \eqref{eq:weight-approximation-obj} to obtain ${\bm \varphi}^\ast$.
\State \Return ${\bm \varphi}^\ast$
\end{algorithmic}
\end{algorithm}

\begin{algorithm}[t]
\caption{Inference with the Proposed \gls{NL}-CMS-\gls{ELM}}
\label{alg:inference}
\begin{algorithmic}[1]
\State \textbf{Inputs:} Inference data point $\mathbf{x}$, \gls{TX}-\gls{RX} shannel $\mathbf{H}$, \gls{MS} propagation coefficients $\{\mathbf{H}_l\}_{l=1}^L$ and $\mathbf{h}_L$, \gls{MS} weights ${\bm \varphi}^\ast$.
\For{$i=1,\dots,L$}
\State Set ${\bm \phi}_l = {\bm \phi}^\ast_l$ to the $l$-th \gls{MS}.
\EndFor
\State Construct transmit signal $\bar{\mathbf{x}}$ via \eqref{eq:transmit-signal}.
\State Transmit $\bar{\mathbf{x}}$ to obtain $\mathbf{y}$ at front \gls{MS} layer via \eqref{eq:received-signal}.
\State Obtain $g(\mathbf{y})$ via the activation of \eqref{eq:activation-function}.
\State Obtain inferred value $\hat{z}$ at the \gls{RX} \gls{RF} chain via \eqref{eq:output-signal}.
\State \Return $\hat{z}$
\end{algorithmic}
\end{algorithm}

\begin{figure*}[t]
    \centering
    \begin{subfigure}[b]{0.32\textwidth}
        \centering
        \includegraphics{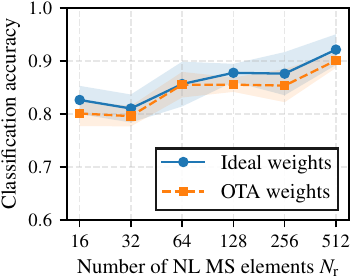}
        \caption{Parkinson's}
        \label{fig:results_parkinsons}
    \end{subfigure}
    \hfill
    \begin{subfigure}[b]{0.32\textwidth}
        \centering
        \includegraphics{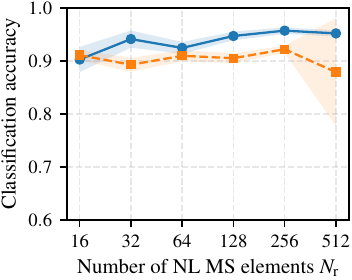}
        \caption{WBCD}
        \label{fig:results_breast_cancer}
    \end{subfigure}
    \hfill
    \begin{subfigure}[b]{0.32\textwidth}
        \centering
        \includegraphics{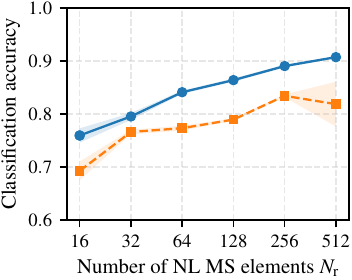}
        \caption{MNIST}
        \label{fig:results_mnist}
    \end{subfigure}
    \caption{Classification accuracy of two \gls{NL}-CMS-\gls{ELM} variations versus the number of elements $N_{\rm r}$ at the \gls{RX}'s front \gls{MS} layer, which corresponds to the number of \gls{ELM} trainable parameters for three distinct datasets. The ``ideal weights'' variation assumes the \gls{ELM} weights are applied in the digital domain or under idealized analog hardware. The ``\gls{OTA} weights'' variation approximates those weights through $L$ cascaded \glspl{MS} with linear responses for full \gls{OTA} inference.}
    \label{fig:overall_results}
\end{figure*}

\subsection{Discussion on Universal Approximation}\label{sec:discussion}

A highly desired property of the proposed \gls{NL}-CMS-\gls{ELM} framework is its capability to approximate arbitrary mappings. Typical \glspl{ELM} are associated with their own versions of the universal approximation theorems~\cite{HuaZhuSie:06, HuaLiCheSie:08}, which mostly consider linear/affine transformations at the hidden layer, followed by activations belonging to families of functions guaranteeing different sets of conditions on non-linearity, boundedness, and/or smoothness. It is thus possible to show that, for some arbitrary value of $N_{\rm r} \leq D$, an \gls{ELM} may approximate the $\mathbf{x} \to y$ mapping with arbitrarily small error.

In the context of \gls{XL} \gls{MIMO} systems, it has been recently demonstrated in~\cite{Stylianopoulos_MIMO_ELM} that, if only the in-phase (or the quadrature) component of the wireless system is utilized for transmission, the universal approximation theorem is still applicable, as long as the channel exhibits Rayleigh-like rich scattering. Under these fading conditions, it has been shown that the \gls{XL} \gls{MIMO} system can be transformed to a standard real-valued \gls{ELM}. In particular, channel diversity is required so that the measurement matrix $\mathbf{G}$ of~\eqref{eq:hidden-layer-outputs} becomes full-rank, and is therefore well-conditioned for inversion when performing \gls{LS} in~\eqref{eq:weights-opt}.
Conceptually, wireless channels (ideally, spatially uncorrelated~\cite{4786505}) act as orthogonal random projections on a feature space, each one extracting a different feature of the data. Thus, the combining operation of the output layer, represented by~\eqref{eq:output-signal}, is trained to learn which linear combination of the random features is associated with each target class.
The exact version of the \gls{NL}-CMS-\gls{ELM} proposed in this paper does not include strictly affine transformation on its hidden layer, due to the fact that the bias term and the NL activation are applied only to the amplitude of $\mathbf{y}$ in~\eqref{eq:activation-function}.
As a result, proving its associated universal approximation theorem necessitates a rigorous and dedicated analytical treatment; this is therefore left as future work~\cite{Our_SPAWC_paper}.
It is noted, nevertheless, that the conditions about rich scattering conditions and large values of $N_{\rm r}$ (i.e., \gls{XL} multi-antenna reception) are still required, regardless of the exact transformation of the hidden layer.

\subsection{Computational Complexity}\label{sec:discussion_compl}

The time complexity for obtaining the solution of~\eqref{eq:LS-objective} is $\Theta(D N_{\rm r} \min\{D, N_{\rm r}\})$, stemming from the matrix inversion of~\eqref{eq:weights-opt}. For reasonably small datasets with \gls{XL} \gls{MIMO} systems where $D = \Theta(N_{\rm r})$, the complexity reduces to $\Theta(N^3_{\rm r})$, which is equivalent to typical \gls{MIMO} decoding operations (such as zero forcing and weighted mean square error) and, therefore, may be computed within a channel coherence frame.
The time complexity of the gradient descent solution of~\eqref{eq:weight-approximation-obj} is $\Theta(T N_{\rm r} L N_{\rm max}^2)$, which can be reduced to $\Theta(T L)$ assuming appropriate parallel hardware, where $T$ is the number of iterations until convergence and $N_{\rm max} \triangleq \max\{ N_1, \dots, N_L\}$.
Considering the assumption of large $N_{\rm r}$ values, the convergence of the gradient is not practically fast enough to be implemented within a single channel coherence block. However, it does not depend on the size of the dataset, which offers a remarkable computational improvement over typical backpropagation-based learning of \glspl{DNN}.
Importantly, the recent work~\cite{Stylianopoulos_MIMO_ELM} demonstrated that, when the channel exhibits correlated fading, only a low complexity re-training of the \gls{XL}-\gls{MIMO}-\gls{ELM} architecture is necessary, since the optimal weights change marginally.
It is therefore feasible to re-tune such models as the channel changes.
We defer the adoption of such techniques for our NL-CMS-ELM framework to the journal version of this work due to a lack of space.

\begin{figure}[t]
    \centering
    \includegraphics[width=\linewidth]{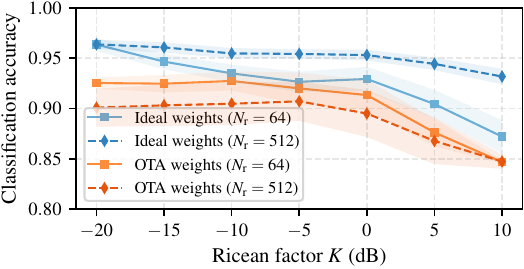}
    \label{fig:ricean_breast_cancer}
    \caption{Classification accuracy of two \gls{NL}-CMS-\gls{ELM} variations over different Ricean factors for the \acrshort{WBCD} dataset, considering different values $N_{\rm r}$ for the \gls{RX}'s front \gls{MS} layer.}
    \label{fig:results-ricean}
\end{figure}

\section{Numerical Results and Discussion}
We perform investigations on the performance of the proposed \gls{NL}-CMS-\gls{ELM} over a number of standard binary classification datasets of small-to-medium size under a variety of system conditions.
The considered datasets~\cite{UCI} are the Parkinson's and the \gls{WBCD} of $22$ and $30$ numerical features for disease diagnosis, respectively, as well as the MNIST dataset of handwritten digit image recognition~\cite{MNIST}.
Due to the large dimensionality of the latter dataset, we have subsampled $100$ pixels from the images at random locations to define the input features. 
To convert this dataset into a binary classification one, we have assumed that the two classes correspond to even and odd digits, respectively.

For pre-processing, all features have been independently scaled to $[0,1]$.
As explained in Section~\ref{sec:system-model}, the input features are encoded in the amplitudes of the transmitted signal vector, following an \gls{AM} transmission scheme.
Consequently, the classification decision based on the output of the system through~\eqref{eq:output-signal} is expressed as $\hat{c} = \mathbbm{1}_{\mathfrak{Re}(\hat{z})>0.5}$ and the classification accuracy, used as our evaluation metric, was measured as $\sum_{i=1}^{|\mathcal{D'}|} \mathbbm{1}_{\mathfrak{Re}(z)^{(i)} = \hat{c}^{(i)}}/|\mathcal{D'}|$ for each of the test sets $\mathcal{D'}$, which have been created via a $70:30$ split in the data points. In our \gls{XL} \gls{MIMO} system, $N_{\rm t}$ is determined by the number of features in the datasets, and we have varied the values of $N_{\rm r}$ to investigate performance scaling. 
Unless otherwise specified, $\mathbf{H} \sim \mathcal{CN}(\mathbf{0}, P_L \mathbf{I})$, where $P_L$ is the pathloss set to $-50$ dB.
The phases ${\bm \vartheta}$ of the \gls{AM} signals were set to $\mathbf{0}$ for all data points.
Similarly, we have sampled $\mathbf{b}$ from a Rayleigh distribution with scale parameter $\mathbb{E}[\|\mathbf{H}\|_{\rm F}]/(2 N_{\rm r} N_{\rm t})$ for the biases to be in the same order as $|\mathbf{y}|$.
Since $z$ contains only $1$ bit of information, we have set the receive \gls{SNR} to a moderate level of $15$ dB in order to assess the performance without severe noise degradation.
A validation under different \gls{SNR} levels and erroneous system information will be included in the journal version. We have used $L=5$ linear \gls{MS} layers, each consisting of $N_l = 64 \times 64$ diffractive elements.
Similarly, we considered $\mathbf{H}_l \sim \mathcal{CN}(\mathbf{0}, P'_L \mathbf{I})$ with $P'_L$ set to $-10$ dB.
Finally, we set the regularization weight $\ell=10^{-6}$ to avoid severe overfitting, and we allowed a maximum of $T=1500$ iterations of the employed \gls{PGD} procedure with step size $0.01$, although convergence was achieved far earlier for most scenarios.

Two main variations of the proposed \gls{NL}-CMS-\gls{ELM} system were investigated. The first one, referred to as ``OTA weights,'' implements Algorithms~\ref{alg:training} and~\ref{alg:inference}, where linear \glspl{MS} are used to approximate $\mathbf{w}^\ast$ \gls{OTA}. We also considered the idealized case where the optimal weight vector is used directly to compute the output as $\hat{z} = (\mathbf{w}^\ast)^\top g(\mathbf{y})$, instead of~\eqref{eq:output-signal}. This approach, referred to as ``ideal weights,'' considers the \gls{RX}'s CMS structure with only the front \gls{NL} \gls{MS} layer (i.e., $L=0$), whose $N_{\rm r}$ outputs are guided to be multiplied with the $\mathbf{w}^\ast$ elements. This idealized weighting can be realized either with digital (requires $N_{\rm r}$ \gls{RF} chains) or analog (a single \gls{RF} chain suffices) combining. For the latter option, phase shifters as in~\cite{Stylianopoulos_MIMO_ELM} or an \gls{MS} structure as in~\cite{shlezinger2019dynamic} with lossless waveguides can be used. Moreover, we report that a $3$-layer fully connected \gls{DNN} trained on the considered datasets achieved $92\%$-$98\%$ classification accuracy, which constitutes an upper bound. The trained \glspl{ELM} with the same number of trainable parameters as the $N_{\rm r}$ values, in every case, achieved similar performance to the \gls{NL}-CMS-\glspl{ELM}, and are thus omitted for clarity.

The performance of the two considered \gls{NL}-CMS-\gls{ELM} variations for an increasing number of elements $N_{\rm r}$ at the \gls{RX}'s front \gls{MS} layer is displayed in Fig.~\ref{fig:overall_results}.
It is noted that the values of $N_{\rm r}$ also represent the number of \gls{ELM} trainable parameters, which affects the capacity of the considered models in solving the classification problems. As observed, in all cases, the classification accuracy increases as $N_{\rm r}$ increases, approaching the upper bounds provided by the theoretical \gls{DNN} benchmarks in the largest \gls{XL} \gls{MIMO} cases. It is also shown that the approximation offered by the linear \gls{MS} layers exhibits close performance to the \gls{ELM} approach using the ideal weights, although, for the largest $N_{\rm r}$ values, a noticeable performance degradation occurs due to insufficient \gls{PGD} convergence.

Finally, we have assumed pure Rayleigh conditions in the \gls{TX}-\gls{RX} channel, implying that $\mathbf{H}$ becomes full-rank (as discussed in Section~\ref{sec:discussion}, these fading conditions are desirable for boosting the performance of the proposed \gls{NL}-CMS-\gls{ELM} framework).
For the results reported in Fig.~\ref{fig:results-ricean}, we have relaxed this assumption to sample channel matrices from a Ricean model~\cite{6184250} in increasing \gls{LoS} conditions, which is mathematically defined as follows:
\begin{equation}\label{eq:ricean}
   \mathbf{H} = \sqrt{P_L} \left( \sqrt{\frac{K}{1+K}} \mathbf{H}_{\rm LoS} + \sqrt{\frac{1}{1+K}} \mathbf{H}_{\rm NLoS} \right),
\end{equation}
where $\mathbf{H}_{\rm LoS}$ is a rank-$1$ matrix of steering vectors, $\mathbf{H}_{\rm NLoS} \sim \mathcal{CN}(\mathbf{0}, 1/\sqrt{N_{\rm t} N_{\rm r}}\mathbf{I})$, and $K$ is the Ricean factor that controls the dominance of either component. It can be observed from the figure that different versions of the proposed \gls{NL}-CMS-\gls{ELM} system have a steady performance when the channel is sufficiently diverse. However, as the channel becomes \gls{LoS}-dominant, the classification accuracy decreases. This is attributed to the fact that, in such cases, fewer independent random features are extracted and $\mathbf{G}$ becomes column-deficient.


\section{Conclusion}
This work showcases that \gls{XL} \gls{MIMO} systems with purposely designed \gls{MS} components at the \gls{RX} side can perform computations that are akin to a complex-valued \gls{ELM} model, and can thus be used to perform \gls{ML}-based inference on the data the \gls{TX} transmits completely \gls{OTA}.
Based on rich scattering conditions, the channel coefficients have been treated as random weights of the \gls{XL}-\gls{MIMO}-\gls{ELM}'s hidden layer, while the \gls{RX}'s CMS, comprising a diffractive \gls{MS} with pseudorandom \gls{NL} response and multiple diffractive \glspl{MS} with trainable linear responses, followed by a single reception \gls{RF} chain has been deployed to provide the activation function and bias directly in the \gls{RF} domain.
The linear \gls{MS} layers were also utilized to approximate \gls{OTA} the optimal trainable \gls{ELM} weights.
A two-step training approach was presented for the proposed \gls{NL}-\acrshort{CMS}-\gls{ELM} approach, along with a discussion on its theoretical performance conditions and guarantees as well as computational complexity. The presented numerical investigation demonstrated classification performance approaching digital \gls{ML} models in the \gls{XL} \gls{MIMO} regime, while resilience to channel diversity levels has been also shown.

\balance

\bibliographystyle{IEEEtran}
\bibliography{Biblio}

\end{document}